\documentclass[12pt]{article}
\usepackage{graphicx}
\begin{document}
\centerline{\bf Comparison of Ising magnet on directed versus undirected}
\centerline{\bf Erd\"os-R\'enyi and scale-free networks}
 
\bigskip
\centerline{M.A. Sumour$^1$, A.H. El-Astal$^1$, F.W.S. Lima$^2$, M.M. Shabat$^3$, and H.M. Khalil$^4$} 
\bigskip
\noindent
$^1$ Physics Department, Al-Aqsa University, P.O.4051, Gaza, Gaza Strip,
Palestinian Authority.\\
$^2$ Departamento de F\'{\i}sica, 
Universidade Federal do Piau\'{\i}, 64002-150, Teresina - PI, Brazil. \\
$^3$ Physics Department, Islamic University, P.O.108, Gaza, Gaza Strip,
Palestinian Authority. Presently at Max Planck Institute for the Physics of 
Complex Systems, N\"othnitzer Stra\"se 38, 01187 Dresden, Germany\\ 
$^4$ Physics Department, Ain Shams University, Cairo, Egypt. \\
\medskip
  e-mail: msumoor@alaqsa.edu.ps (MS), wel@ufpi.br (FWSL)

\bigskip

{\bf Abstract}

Scale-free networks are a recently developed approach to model the interactions 
found in complex natural and man-made systems. Such networks exhibit a 
power-law distribution of node link (degree) frequencies $n(k)$ in which a 
small number of highly connected nodes predominate over a much greater number 
of sparsely connected ones. In contrast, in an Erd\"os-R\'enyi network 
each of $N$ sites is connected to every site with a low probability $p$ (of the order of $1/N$). Then the number $k$
of neighbors will fluctuate according to a Poisson distribution. One can instead
assume that each site selects exactly $k$ neighbors among the other sites. 
Here we compare in both cases the usual network with the directed network, when site A selects site B as a neighbor, and then B influences A but A does not 
influence B. As we change from undirected to directed scale-free networks, the 
spontaneous magnetization vanishes after an equilibration time following an 
Arrhenius law, while the directed ER networks have a positive Curie temperature.


\bigskip
{\bf \large Introduction}
 
\bigskip
This paper deals with Ising spin on (mostly) directed Erd\"os-R\'enyi (ER) random
graphs \cite{reny1, reny2, reny3} and Barab\'asi-Albert (BA) scale free networks
\cite{ba}. Sumour and Shabat \cite{sumour,sumourss} investigated Ising models with
 spin $S=1/2$ on directed BA networks \cite{ba} with
 the usual Glauber dynamics.  No spontaneous magnetisation was 
 found, in contrast to the case of undirected  BA networks
 \cite{alex,indekeu,bianconi} where a spontaneous magnetisation was
 found below a critical temperature which increases logarithmically with
 system size. In $S=1/2$ systems on undirected, scale-free hierarchical-lattice
 small-world networks \cite{nihat}, conventional and algebraic
 (Berezinskii-Kosterlitz-Thouless) ordering, with finite transition
 temperatures, have been found. Lima and Stauffer \cite{lima} simulated
 directed square, cubic and hypercubic lattices in two to five dimensions
 with heat bath dynamics in order to separate the network effects  form
 the effects of directedness. They also compared different spin flip
 algorithms, including cluster flips \cite{wang}, for
 Ising-BA networks. They found a freezing-in of the 
 magnetisation similar to  \cite{sumour,sumourss}, following an Arrhenius
 law at least in low dimensions. This lack of a spontaneous magnetisation
 (in the usual sense)
 is consistent with the fact
 that if on a directed lattice a spin $S_j$ influences spin $S_i$, then
 spin $S_i$ in turn does not influence $S_j$, 
 and there may be no well-defined total energy. Thus, they show that for
 the same  scale-free networks, different algorithms give different
 results. The $q$-state Potts model has been studied in scale-free networks
 by Igloi and Turban \cite{igloi} and depending on the value of $q$ and the
 degree-exponent  $\gamma$ first- and second-order phase transitions
 are found, and also by Lima \cite{lima2} on directed 
 BA network, where only first-order phase transitions
 have being obtained independent of values of $q$ for values of 
 connectivity $z=2$ and $z=7$ of the directed BA network.
 More recently, Lima \cite{lima1} simulated the Ising model
 for spin $S=1$ on directed 
 BA network and different from the Ising model for
 spin $S=1/2$, an {\it unusual} order-disorder phase transition of 
 order parameter was seen; this effect needs to be re-evaluated in the light of 
 the time dependence presented below.  

In previous work \cite{sumour,sumourss} we also studied
 the Ising model on directed BA networks, by checking the magnetization on it. Analogously in this work we check the modified ER network (or Wilf graph) by  taking
an exact number of neighbors like in BA network, then we work in directed ER 
network with low probability, and finally go to undirected ER network. 
We also study the Ising model for spin $S=1/2$, 1, 3/2 and 2 on directed 
BA network. In all cases we check whether or not a spontaneous
magnetization exisst in equilibrium. The Ising model with spin 1/2 on the directed
ER graphs and that with spin $S=1/2$, 1, 3/2 and 2 on BA networks
was seen not to show  a {\it usual} spontaneous 
 magnetisation and this decay time for flipping of the magnetisation
 followed an  Arrhenius law for HeatBath algorithms that agrees with the
 results of the Ising model for
 spin $S=1/2$ \cite{sumour,sumourss} on directed BA
 network.

\bigskip

{\bf \large Model and Simulation}

\begin{figure}[hbt]
\begin{center}
\includegraphics [angle=-90,scale=0.33]{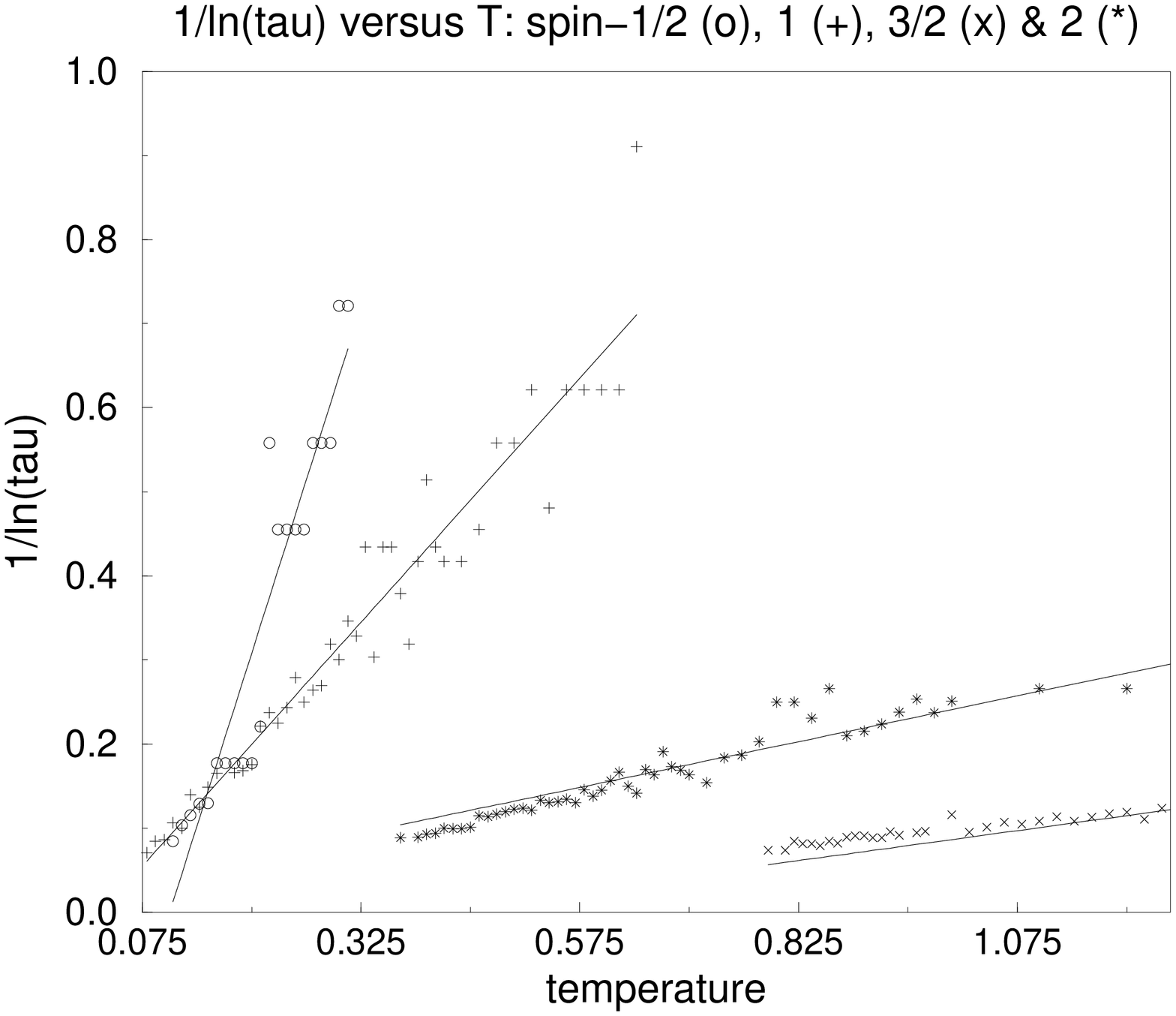}
\includegraphics [angle=-90,scale=0.33]{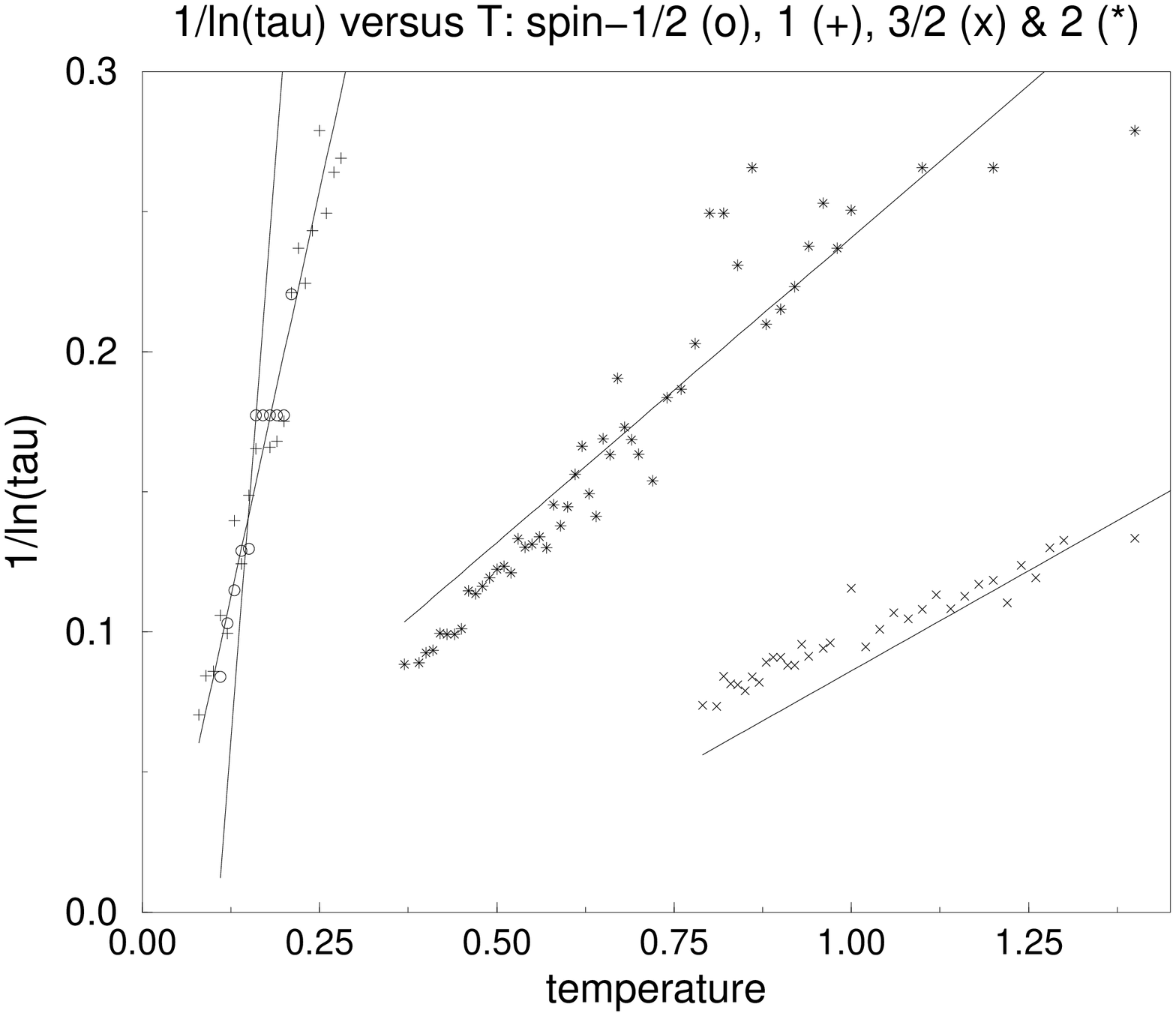}
\end{center}
\caption{Reciprocal logarithm of the relaxation times on directed BA networks
for $S = 1/2$ to $S = 2$. The right part is a zoom of the left part for the
longest times.}
\end{figure}

\begin{figure}[hbt]
\begin{center}
\includegraphics [angle=-90,scale=0.5]{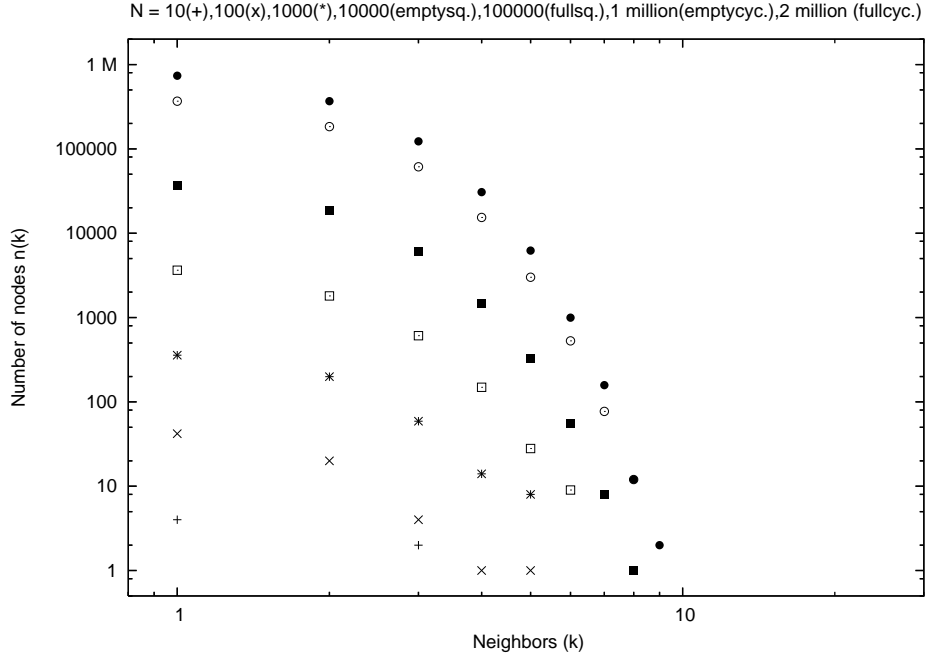}
\end{center}
\caption{$n(k)$ versus $k$ with different $N$ (10, 100, 1000, 10000, 100000), 
for exactly 4 selected neighbors.}
\end{figure}

\centerline{\bf Ising model on directed Barab\'asi-Albert Networks}

We consider the spins $S=1/2$, 1, 3/2 and 2 Ising model on directed
Barab\'asi-Albert (BA) networks, defined by a set of
spin variables taking the values $\pm 1$ for $S=1/2$, $\pm 1$ and
0 for $S=1$, $\pm 3/2$ and $\pm 1/2$ for $S=3/2$,  and $\pm 2$, $\pm 1$ and
0 for $2$, respectively, situated on every site of a directed
BA networks with $N$ sites.

The probability for spin $S_{i}$ to change its state in these directed  networks is
\begin{equation}
p_{i}= 1/[1+ \exp(2E_{i})/k_BT)], \quad  E_{i}=-J\sum_{j}S_{i}S_{j}
\end{equation}
where the $j$-sum runs over all selected neighbors of $S_{i}$. In this network,
each new site added to the network selects (preferential attachment proportional
to the number of previous selections) with connectivity $z=2$
already existing sites as neighbours influencing it; the newly
added spin does not influence these neighbours.

To study the spins $S=1/2$, $1$, $3/2$ and $2$ Ising model we start with
all spins up, a number of spins equal to $N=500000$, and time up $2,000,000$ 
(in units of Monte Carlo steps per spin), with HeatBath Monte Carlo algorithm. 
Then we vary the temperature $T$ and at each $T$ study the time dependence for  
9 samples. The temperature is measured in units of critical temperature of the
square-lattice Ising model. We determine the time $\tau$ after which the
magnetisation has flipped its sign for the first time, and then take the
median values of our nine samples. So we get different values $\tau_{1}$
for different temperatures.

In this study of the critical behavior this Ising model (with spins $S=1/2$,
1, 3/2 and 2) we define the variable $m=\sum_{i=1}^{N}S_{i}/N$ as  
normalized magnetization.

Our BA simulations, using the HeatBath algorithm, indicate that the spin
$S=1/2$, 1, 3/2 and 2 Ising model does not display a phase transition and 
the plot of the time 
$1/\ln\tau$ versus temperature in Fig. 1 shows that our BA results for all spins
agree with the modified Arrhenius law for relaxation time, defined as the
first time when the sign of the 
magnetisation flips: $1/\ln(\tau) \propto T + \dots$.

\bigskip
\centerline{\bf Modified ER network}

In the classical ER model all edges are equally probable and independent.
We take a modification where each node connects with an exact number of neighbors 
like in the BA network; we take number of neighbors as 4. So we plot the number 
$n(k)$ of nodes versus  the number $k$ of neighbors in Fig. 2.

Fig.2 does not have the shape of the corresponding Poisson distribution for $ <k> 
= 4$. 
The maximum number of nodes which select a site as neighbor is seen in Fig.3
to vary roughly logarithmically with the size $N$ of the network.

\bigskip
 
\begin{figure}[hbt]
\begin{center}
\includegraphics [angle=-90,scale=0.5]{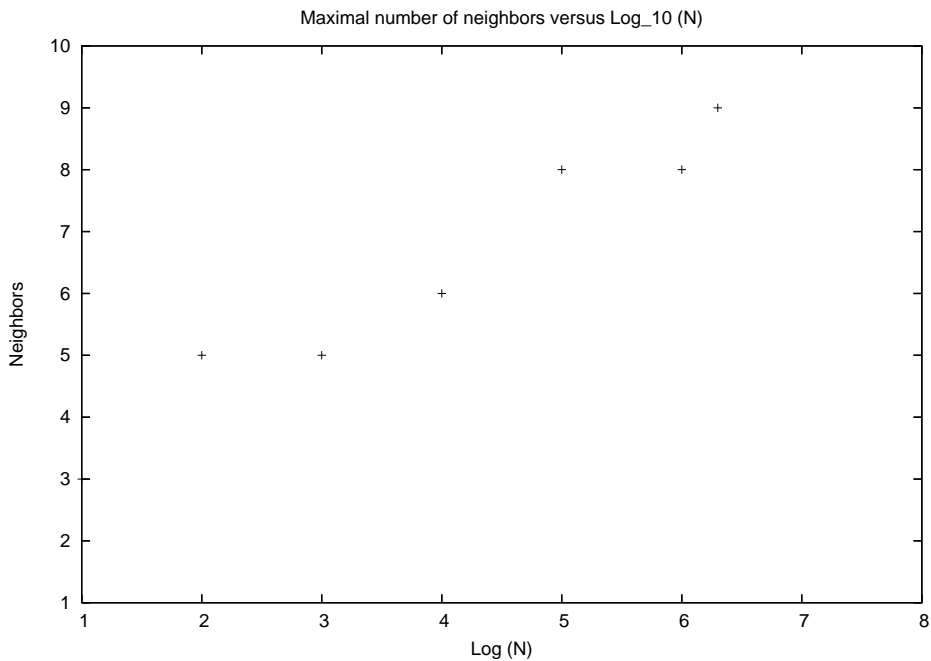}

\end{center}
\caption{Maximum number of neighbors versus decadic logarithm of the 
number $N$ of nodes, for 4 neighbor selections per node.}
\end{figure}

\bigskip

\centerline{\bf Ising model on modified Erd\"os-R\'enyi network}

When we put the Ising model on the Erd\"os-R\'enyi network with size of network 
= 2 million , temperatures 0.184 to 0.310, number of neighbors = 2 and 4, and 
time = 20000, and observe the time when the magnetization starts to change it's 
sign, we get Fig.4.

\begin{figure}[hbt]
\begin{center}
\includegraphics [angle=-90,scale=0.3]{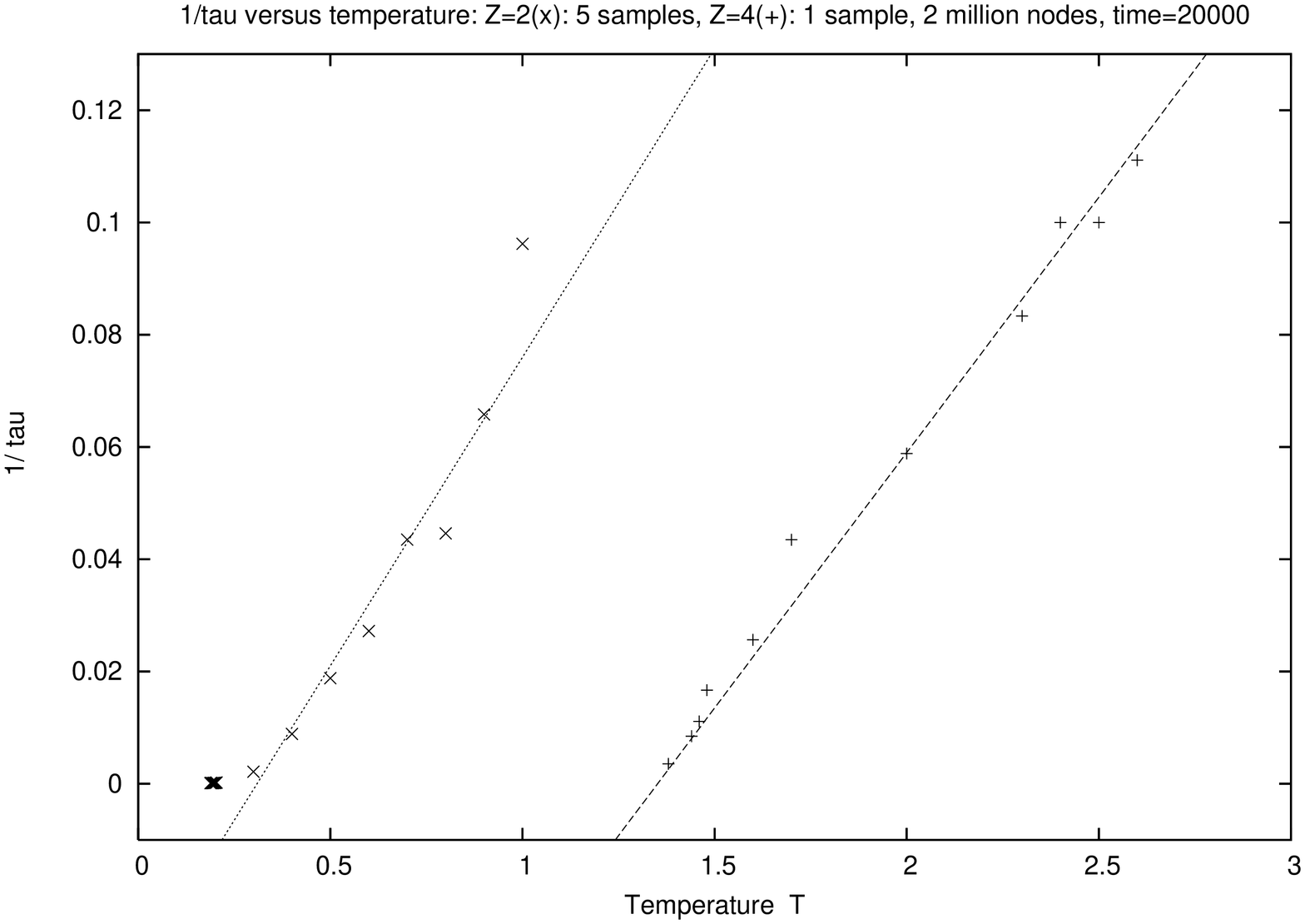}
\includegraphics [angle=-90,scale=0.3]{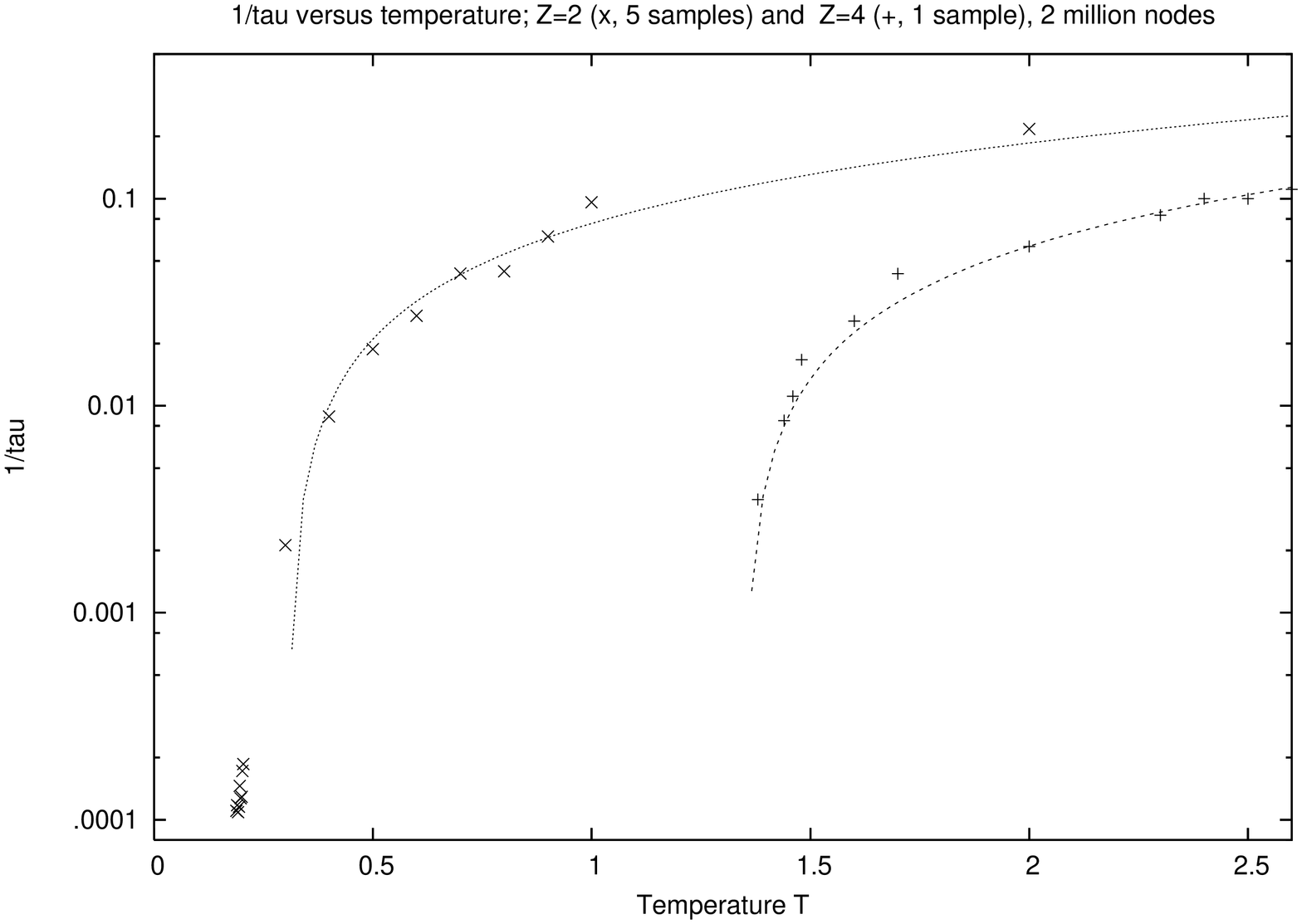}
\end{center}
\caption{Reciprocal relaxation time $1/\tau$ versus temperature with network 
size = 2 and 4 million, time = 20000, for modified ER with exactly 2 or 
4 neighbors. The upper part uses a linear, the lower a logarithmic scale for 
the reciprocal relaxation time.}
\end{figure}

\bigskip

The reciprocal relaxation time varies linearly with temperature 
for $z = 2$ and 4 neighbors selected by each new node, and can be extrapolated
to vanish at some positive Curie temperature $T_c$: The relaxation time goes to
infinity there. 

\bigskip

\centerline{\bf Erd\"os-R\'enyi (ER) Network}

Now we move from the modified to the classical ER graphs.
In the normal ER  model all edges are equally probable and appear independently.
To get the proper  number of nodes of normal ER network  we need $ N \gg 1$ and 
probability $p \ll 1$ with $pN$ of order unity; each edge is chosen to appear 
with probability $p$. We used $p = 1/N$, $p = 2/N$, and $p=3/N$. 
On average, each vertex will have a {\it small} number of neighbors.
We found no significant difference between the two networks for large size in 
Poisson distribution and observed degree distribution $n(k)$.

\bigskip
\centerline{\bf Ising Model on Directed ER Network}

We take different probabilities for different number of nodes $N$ (1000, 10000, 
50000, 100000), with different temperatures in Fig.5.  There we check again the
first time after which the magnetization changes sign, 
take the median from nine samples, and plot from the reciprocal of the time for
three probabilities ($p=3/N, \; p=2/N, \, p=1/N$) in Fig.5.

\bigskip

\begin{figure}[hbt]
\begin{center}
\includegraphics [angle=-90,scale=0.5]{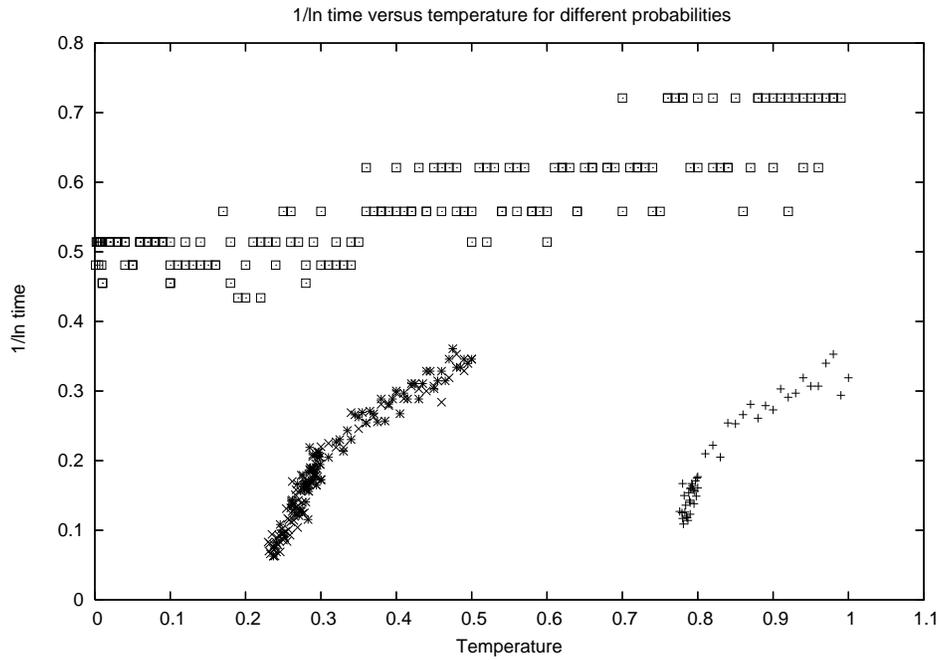}

\end{center}
\caption{ 1/ln(time) versus temperature for different probabilities $1/N$ (sq.),
$2/N$ (x), $3/N$ (+).}
\end{figure}

\begin{figure}[hbt]
\begin{center}
\includegraphics [angle=-90,scale=0.5]{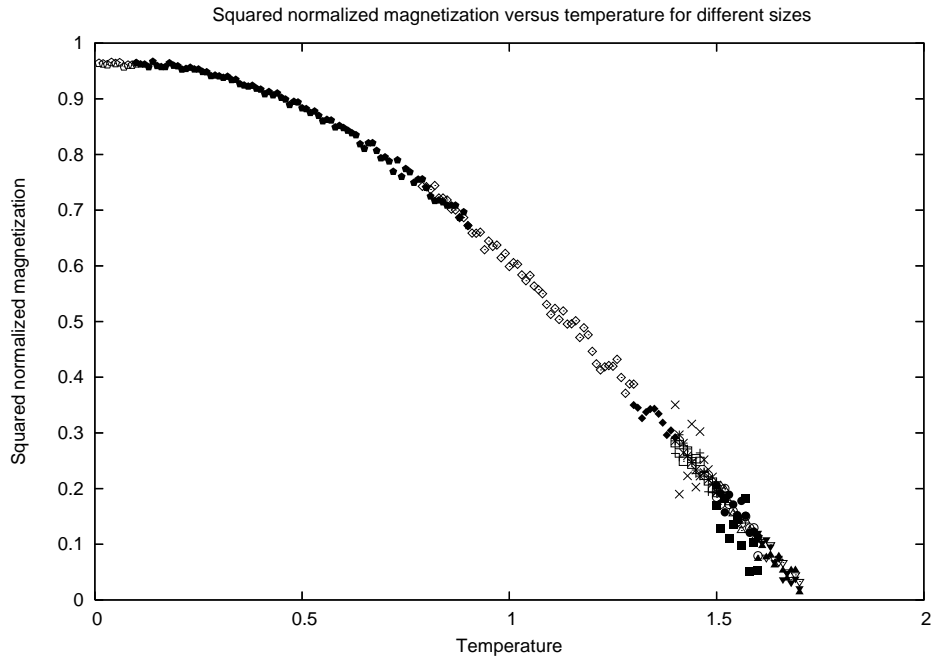}
\end{center}
\caption{ Squared normalized magnetization versus temperatures for different sizes $N$ of the undirected ER graph.}
\end{figure}

The figure shows nicely the difference between probability $1/N$ ( = percolation
threshold) and larger probabilities. This figure shows that there is a spontaneous magnetization at $p=2/N$ for the left curve and the right curve for $p=3/N$, but not a spontaneous magnetization at $p=1/N$ which is the percolation threshold.

\bigskip
\centerline{\bf Ising model on undirected ER network}

For the undirected network we use only one probability equal $p = 2/N$, 
because it gives a clear answer compatible with the mean-field universality
class, as expected because of the infinite range of the symmetric interaction. 
Before, for directed networks or graphs, when a new site A selects on old site 
B as a neighbor, then we had only one direction. Now, for undirected network, 
not only A is a neighbors of B but also B is a neighbor of A. 
From our simulation we see that  the undirected version has a spontaneous magnetization, to which the system relaxes similarly to the standard Ising square 
lattice, 
Then we plot the square of normalized magnetization versus temperature in Fig.6.
In equilibrium there is a Curie temperature $T_c$: below $T_c$ we have a 
spontaneous magnetization and above $T_c$ we do not have one as we see in Fig 6.
The squared magnetization vanishes at this $T_c \simeq 1.7 J/K_B$ linearly in 
temperature.
This behavior corresponds, not unexpectedly, to a mean field critical exponent.

\bigskip
{\bf \large Discussion}

For the directed BA networks we found an Arrhenius law. This means that for each 
positive temperature there is a finite relaxation time after which the
magnetisation decay towards zero. Similarly to the one-dimensional Ising
model there is no ferromagnetism on this directed Barab\'asi-Albert
network.

A directed (normal or modified) ER network has a phase transition temperature 
below which a spontaneous magnetization exists, while the directed BA network 
has no such phase transition. The undirected ER network has a spontaneous 
magnetization in the universality class of mean field theory. 

  We thank D. Stauffer for many suggestions and fruitful
discussions during the development this work and also for the revision of
this paper. Lima also acknowledges the Brazilian agency FAPEPI
(Teresina-Piau\'{\i}-Brasil) for  its financial support and
the computational park CENAPAD.UNICAMP-USP, SP-BRASIL.
for supporting the system SGI Altix 1350.

\end{document}